\documentclass[conference]{IEEEtran}
\IEEEoverridecommandlockouts
\usepackage{cite}
\usepackage{amsmath,amssymb,amsfonts}
\usepackage{algorithmic}
\usepackage{graphicx}
\usepackage{textcomp}
\usepackage{xcolor}
\usepackage{verbatim}

\def\BibTeX{{\rm B\kern-.05em{\sc i\kern-.025em b}\kern-.08em
    T\kern-.1667em\lower.7ex\hbox{E}\kern-.125emX}}

\usepackage{todonotes}

\begin{document}

\title{Test Automation Process Improvement in a DevOps Team:  Experience Report  \\
{\footnotesize \textsuperscript{}}
\thanks{$^{1}$ Contribute equally to this study}
\thanks{$^{*}$ Corresponding author}
\thanks{\copyright~2020 IEEE . Personal use of this material is
permitted. Permission from IEEE must be obtained for all other uses, in any current or future media, including reprinting/republishing this material for advertising or promotional purposes, creating new collective works, for resale or redistribution to servers or lists, or reuse of any copyrighted component of this work in other works.}
}

\author{\IEEEauthorblockN{1\textsuperscript{st} Yuqing Wang$^{1 *}$}
\IEEEauthorblockA{\textit{M3S research unit, University of Oulu} \\
Oulu, Finland \\
yuqing.wang@oulu.fi}
\and
\IEEEauthorblockN{2\textsuperscript{nd} Maaret Pyh{\"a}j{\"a}rvi$^{1*}$}
\IEEEauthorblockA{\textit{F-Secure} \\
Helsinki, Finland\\
maaret.pyhajarvi@f-secure.com}
\and
\IEEEauthorblockN{3\textsuperscript{rd} Mika V. M{\"a}ntyl{\"a}}
\IEEEauthorblockA{\textit{ M3S research unit, University of Oulu} \\
Oulu, Finland \\
mika.mantyla@oulu.fi}
}

\maketitle

\begin{abstract}
How to successfully conduct test automation process improvement (TAPI) for continuous development, consisting of iterative software development, continuous testing, and delivery, is the challenge faced by many software organizations. In this paper,  we present an experience report on TAPI in one DevOps team in  F-Secure (a  Finnish software company). The team builds  Windows application software and exists in  F-Secure’s TAPI  culture. The team self-reports high satisfaction and maturity in test automation for continuous development. To study their TAPI, we reviewed a collection of experience notes, team reflection reports and telemetry result reports. Then several meetings were held to discuss the details. We found that based on the understanding of the team, test automation maturity for continuous development is defined as a set of indicators, e.g., the increasing speed to release, improving the productivity of the team, high test efficiency. Second, the team indicated that a set of critical success factors have a major impact on successfully carrying out its TAPI, e.g., incremental approach, the whole team effort, test tool choice and architecture, telemetry. Third, we compare the TAPI practices in the observed team with the practices described in prior literature. The team believes that the existing test automation maturity approaches should include the identified practices like the whole team effort to build a more comprehensive test automation improvement model for the software industry.

\end{abstract}


\section{Introduction}
\label{sec:introduction}
Nowadays, software organizations bear the pressure to get their products to the market quickly and deploy them frequently~\cite{kroll2018}. Test automation has been widely applied to ensure consistent product quality in the frequent release cycles. However, many organizations still have immature test automation processes with process-related issues such as inefficiency of test activities, heavy maintenance effort, slow feedback to the development work~\cite{rafi2012benefits,wiklund2012}.


According to the state of testing report 2019~\cite{PractiTest2019}, the software industry has been more and more concerned about test automation process improvement (TAPI). Many software organizations are conducting TAPI aimed at achieving for continuous development~\cite{eldh2014,furtado2014}. Continuous development can be seen as 
``process for iterative software development and is an umbrella over several other processes including continuous integration, continuous testing, continuous delivery and continuous deployment"~\cite{Gnew}. However, based on many sources~\cite{kasurinen2010,worldqualityreport18,ISTQB2018}, despite the effort, not all software organizations are able to meet the purpose of TAPI, usually caused by inadequate implementation. 


Several researchers have stated the importance of  TAPI research to increase the likelihood of succeeding with TAPI~\cite{wang2019,wang2018test}. Since TAPI is a new trend in recent years, little empirical research has been conducted on observing software organizations that have been successfully carrying out TAPI and gaining  test automation maturity for continuous development~\cite{garousi2017,hrabovska2019}.  

The purpose of this paper is to present an experience report on TAPI in one DevOps team at F-Secure (a Finnish software company).  The team builds Windows application software and exists in F-Secure's TAPI culture. The team self-reports high satisfaction and maturity in test automation for continuous development. This paper aims to answer the research question:


\newcommand{\RQoneTitle}{RQ1 - Perceived success factors}
\newcommand{\RQone}{What makes TAPI successful in the DevOps team at F-Secure?}

\begin{itemize}
    \item \textbf{\RQoneTitle}: \RQone
\end{itemize}

The second author of this paper is an engineering manager who monitored TAPI in the observed team at F-Secure. She provided a collection of experience notes, team reflection reports, and telemetry result reports. To answer the research question, we reviewed those materials and hold several meetings to discuss the details. The first author conducted thematic analysis  to identify critical factors that make TAPI successful in the team on the available materials. Those factors were verified and revised by the second author to ensure the accuracy and correctness. The detailed study results can be found in the remainder of the paper.




This paper is structured as follows. Section~\ref{sec:background} introduces the background and related work that indicate the reason for our research. Section~\ref{sec:method} introduces the research method and process. Section~\ref{sec:results} presents the study results. Section~\ref{sec:dicussion} discusses the implications to study results and threats to validity.  Section~\ref{sec:conclusion} concludes the study and illustrates the future work.

\section{background}
\label{sec:background}
This section reviews the concept of software test automation and related work conducted on TAPI research.

\subsection{Software test automation}
Test automation is the use of tools (normally referred as test tools) to automatically test the applications in software development~\cite{pocatilu2002}. Similar to general purpose of software development, test automation follows a lifecycle which determines how it begins, evolves,
and ends~\cite{pocatilu2002,garousi2017test}. Test automation covers the entire software test process and consists of the variety of testing activities, e.g., test case design, test scripting, test execution, test evaluation, to test-result reporting~\cite{Vahid2016When}. 


\subsection{Related work}
TAPI related topics have been studied by software engineering (SE) practitioners and researchers for many years.  Many practitioners have published their TAPI related articles in blogs, magazines, SE related websites. For example, there are such articles explain why and how TAPI should be carried out for continuous development~\cite{Tatu}. Other studies (e.g.,~\cite{G1,G2}) that summarize the benefits, challenges, general steps, success factors of TAPI also exist.  

For TAPI research, based on our review on related work, there are survey studies (e.g., \cite{karhu2009,kasurinen2010}) exploring the state of art of test automaton in the industry, and indicates the need for TAPI for many organizations, especially for those who are far from being mature. Several empirical studies (e.g., \cite{pettichord1999,graham2012}) have been explored the steps and practices of conducting TAPI in software organizations. Additionally, there is the number of recent studies conducted test automation maturity models for providing the guidelines for TAPI, for example, Eldh \textit{et al.}~\cite{eldh2014} develops the TAIM model and Furtado \textit{et al.}~\cite{furtado2014} develops MPTA.BR. 

However, despite many TAPI topics have been examined by prior work,  we failed to identify  empirical research conducted on observing software organizations that are carrying TAPI in practices. 

\section{Research method}
\label{sec:method}

We studied several industrial experience reports conducted in software engineering field. We conducted our study in three stages: (1) a study plan, (2) data collection, (3) data analysis. Each stage is described in the following sub-sections.

\subsection{A study plan}
 We defined a study plan to set data collection and analysis process for presenting an experience report. This plan was agreed by all authors.  

\subsection{Data collection}

The second author was
responsible for facilitating the access to the required data
from the team of F-Secure for this study. The data
collection was carried out in two steps.

 In the first step, a collection of experience notes, team reflection reports, software production snapshot data were shared with all co-authors. Experience notes describe the evolution of TAPI practices from 2005 to 2019. They were created by the second author of this paper for purposes of this research. Software production snapshot data from 2019 was collected for sharing telemetry statistics, release and test automation change log excerpts. The first and third authors of this paper studied those materials to familiarize with the case and identify data answering research questions.  
 
In the second step, six online meetings among co-authors were conducted via Skype. The duration was 55-67 minutes. Before a meeting, a meeting guide was prepared to outline the discussion topics including, for example, when, why, and how they do particular TAPI practices (recorded on the materials), and what effect those TAPI practices have. During a meeting, we carried out the discussion around those topics. The second author took responsibility to explain the contents on company experience materials and complement with details. Open questions were asked for answering our research question. The notes were written down at a meeting. All meetings were audio recorded. The audio records of meetings were transcribed verbatim into text files. 

\subsection{Data analysis}
\begin{figure} [h]
\centerline{\includegraphics[width=1\linewidth]{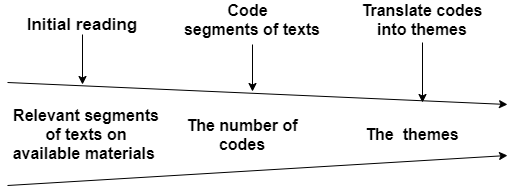}}
\caption{Data thematic analysis process (modified according to~\cite{cruzes2011})}
\label{fig:dataAnalysis}
\end{figure}

The text files that record the transcription verbatim of meetings, meeting notes, as well as experience notes were imported into NVivo (a qualitative data analysis software)~\cite{Nvivo}. We identified critical success factors from those materials by performing inductive coding~\cite{cruzes2011}. Inductive coding is an thematic analysis technique. It uses a iterative approach to extract data from sources and then build the common themes to classify them. Our inductive coding process was performed in three steps, as shown in Fig.~\ref{fig:dataAnalysis}.  

First, the initial reading was performed to identify relevant texts (that describe relevant critical success factors) on available materials. Second, we coded the segment of relevant texts in NVivo. The memos were written throughout the coding process. Third, by reviewing the codes and memos, we establish themes to describe the critical success factors. 

We reviewed the trustworthiness of our list of success factors mapped into categories, when necessary, the original texts were examined.


Based on the thematic analysis, the first author summarized study results to answer the research question in this paper. The second author reviewed the study results and proposed the changes: (1) revise inappropriate contents, (2) complement more examples and details, (3) add new contents considered to be important. The final modification were made in the discussion among co-authors. The final results was reported by the first author and the second author by the paired-simultaneous writing. The third author reviewed that. 

\section{Results}
\label{sec:results}
To present the study results, we first provide an overview of case description. Next,
the research question is answered.

\subsection{The company and team description}
\label{sec:casedescription}
F-Secure Oyj is a cyber security company with headquarters in Finland and over 1600 employees globally. F-Secure's products protect enterprises and consumers against a wide variety of security threats. Windows endpoint security products (incl. features such as antivirus) form a product line sharing code assets and practices. F-Secure has multiple DevOps teams attending to different customer segment's needs, creating different Windows endpoint security products from a common product line.  

With 30 years of history in testing Windows endpoint security products, the start of serious test automation efforts date back to 1999 with a training by Mark Fewster. Efforts lead to tool selection and implementation in place in 2005, following best practices of the time.  Since then, the TAPI initiative was started. 

Nowadays, test  automation  is  an integral  part of  fast-paced  development  giving  developers  feedback at F-Secure for Windows endpoint security products. TAPI is embedded in the culture for continuous development. The goal of TAPI is to enhance the ability to produce and maintain the quality of products in agile and continuous integration (CI) environment. TAPI practices may be different from team to team, but each team shares the TAPI culture at F-Secure.

The DevOps team observed in this paper is Epics, with currently 11 engineers. Epics is responsible for developing and operating Windows endpoint security products that integrate with cloud-based Protection Service for Businesses management system for corporate customers. Epics was created in 2016 to build on an existing consumer product and further develop it for corporate customers. Epics product responsibility has grown from original 1 product to currently 14 products it releases versions on with a monthly schedule. Products Epics operate count their users in millions. Epics shares F-Secure's culture of TAPI and leads efforts in speeding up Windows endpoint security product's release cadence. 

Looking back to 2005, test automation in the similar team operating for Windows endpoint security products looks different. While application development was responsibility of several feature teams, test automation was responsibility of two test automation developers in a separate team providing reusable test libraries as a service for those feature teams. Using a commercial Capture and playback tool as programming platform, the two test automation developers created a library of tests feature teams would run. Separating the creation and maintenance from running created a few particular issues. The coverage of test automation was limited, running it and analyzing results was manual work, skills for maintenance unavailable in feature teams and selected market leader tool evolved out of its market position and was left behind better tools. 

The current generation of test automation efforts with  Epics are founded on a Windows endpoint protection platform, multi-team effort started in 2009 for consumer products and built from whole-team responsibility with open source tools. Both the new product architecture and the test automation system were created as a pair to support one another. Epics joined in 2016 to develop corporate products with the same platform, followed by a second corporate team with different product responsibility in 2018.

\begin{figure}[h]
\centerline{\includegraphics[width=1\linewidth]{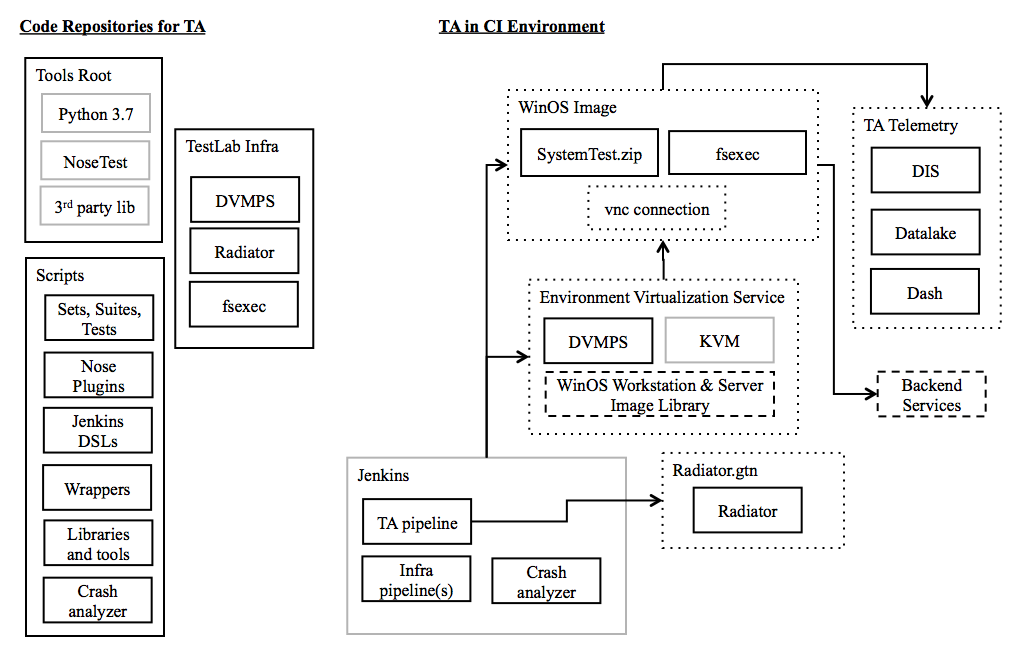}}
\caption{Test automation system}
\label{fig:architecture}
\end{figure}

Fig.~\ref{fig:architecture} shows the current test automation system shared among teams (including Epics) for Windows endpoint products. The whole system consists of seven areas: Tools Root, Scripts, TestLab infra, WinOS image, Environment virtualization service, Jenkins, Radiator and TA telemetry. CI to build test automation system exists and it pulls latest from code repositories for test automation to run test automation in CI environment on product change. 

At the time of writing this paper, Epics self-reports high  satisfaction and maturity in test automation for  continuous development, which is demonstrated by the set of indicators:
 

 \begin{itemize}
    \item \textbf{The increasing speed to release:} Epics has the ability to make continuous release decisions based on test automation results. While regular release cycle has been 1-2 major, 2 minor releases a year for each product team, Epics had 9 releases of its products in 2019 shown in  Table~\ref{tab:releaseTA}. Each release contained hundreds of changes on product and test automation. Time from release decision to release at first customer machines has improved from 5 days (2018) to 4 hours (2019) and the team makes progress towards two week cadence typical for web applications for a Windows application with distributed deployment. 
    
    
    \item \textbf{Improving productivity of the team:} Epics with 11 people was capable splitting their effort to test automation, customer valuable features, maintaining, monitoring, and operating for a large, significantly growing, user base. Epics contributed 2917 code changes (including test automation) to Windows endpoint security platform to take forward products they are responsible for.  
    
    \item \textbf{Shared platform work efficiency:} Quality of the Windows endpoint security platform remained high showing up in low number of maintenance issues, while it was developed actively in multi-team multi-site independent teams. Test automation showed what part of the system in CI is unavailable and when it returns to availability. 
     
    \item \textbf{Sustainable test automation maintenance effort:} Test automation stayed up to speed with changes while team had bandwidth to other work. Every team member contributed to test automation, sharing the load. 
    
    \item \textbf{Finding relevant issues:} Test automation helped add a set of Server products in a few days finding operating system specific issues, identified large number of crashes and pinpointed relevant problems.
    
    \item \textbf{The high satisfaction of customers:} Epics addressed 167 support issues from customer base counted in millions and worked on one support escalation for their products. 
    
    \item  \textbf{High test efficiency:} Run a maximum of 213 708 tests on single working day to cover the changes of that day.
    
    \item \textbf{Reasonable investment for TAPI:} No visible investment on improving test automation process as it is part of normal work. 
\end{itemize}

\begin{table}[htbp]
\caption{Release information in 2019}
\begin{center}
\begin{tabular}{p{1.1cm} p{2cm} p{2cm} p{2cm}}
\hline

\textbf{Release} &\textbf{ \# of commits on product} & \textbf{\# of commits on test automation} & \textbf{Availability} \\
\hline
19.1 & 655 & 398 & 23.01.2019 \\
19.2 & 689 & 298 & 05.03.2019 \\
19.3 & 519 & 349 & 03.05.2019 \\
19.4 & 517 & 255 & 06.06.2019 \\
19.5 & 304 & 184 & 27.06.2019
\\
19.6& 285 & 195 & 12.08.2019
\\
19.7 & 290 & 137 & 05.09.2019
\\
19.8 & 530 & 311 & 28.10.2019 \\
19.9 & 304 & 365 & 19.11.2019
\\
\hline

\end{tabular}
\label{tab1}
\label{tab:releaseTA}
\end{center}
\end{table}


 
 


\subsection{\RQoneTitle}
To answer `\RQoneTitle', we identified critical factors that determine the success of TAPI in the DevOps team at F-Secure. Those factors were classified into several dimensions, see Table~\ref{tab:factors}. In next sub-sections, we elaborate the details about what test automation practices are performed around those factors and what impacts each factor has. 

\begin{table}[htbp]
\caption{Perceived Success Factors}

\begin{center}
\begin{tabular}{p{2cm} p{5.4cm}}
\hline

\textbf{Dimension} &\textbf{Factor}  \\
\hline
Human &  Whole team
effort 
\newline Expert team members \newline Self-motivated team members
\\

\hline

Organizing & 
Allow time for learning curve
\newline Internal open source community mindset 
\\

\hline

Technical & Test tool choice and architecture \newline 
Testlab infrastructure \newline
Product testability\newline
Telemetry\\
\hline

Process & Incremental approach\newline 
Process observation and optimization
\\ \hline

\end{tabular}
\end{center}
\label{tab:factors}
\end{table}

\subsubsection{Whole team effort} The whole team effort is considered as a critical factor for the success of TAPI in Epics.

In 2005, when the two-person in the separate team are creating a library of automated tests for the feature teams for Windows endpoint products, the test automation was isolated from other development work. Maintenance was hindered by lack of test automation skills in feature teams, which were trying to use the created automation. 

In 2009, when product architecture was completely revamped and new test automation created side by side to it, test automation became a developer specialty creating individual Python developers building test automation systems from within the team. 

In 2016, when Epics started working, their practices and tooling for test automation came with the Endpoint protection platform. New hires included a test automation specialist (a new Python developer). Allowing growing to learn Python on the job with continuous progress in TAPI gradually took the team towards supporting continuous testing and frequent releases. 

In 2018, the team moved organically to a `whole team effort model'- \textit{everyone working toward the same goals to build software products} - while serving growing user base and more frequent releases. The goal was to release more often, which required removing possible bottlenecks in test automation. In this model, everyone was encouraged to conduct test automation. Establishing the release practice and seeing each member contributes to test automation activities took a year. 

At the time of writing this paper, the team is cross-functional and carry out agile and CI practices. There are two dedicated specialists in the area of test automation. All team members perform test automation tasks. The team takes care of feature and test automation discovery, development and maintenance, and operate and monitor the production environment. Each member worked with more responsibility to create and enhance the value of automated tests.



\subsubsection{Expert team members} 

One important factor determining success of TAPI is in the mix of people in Epics. From first senior test automation developer who had never written Python to senior test strategist, to senior developers experienced with test automation, to a junior aged 15 when starting with the team, they provided a mix of perspectives and forced deliberate learning as part of the work. Test automation expertise was not test automation developer specialty, but something every team member had perspectives to contribute on. Significant contributions to better test automation came from a continued series of insights implemented in code, e.g., a team interaction leading a Python developer implement the telemetry plugin.

\subsubsection{Self-motivated team members} Self-motivated team members was understood as one major factor. When moving to `the whole team effort' model in Epics, the goal of TAPI was modeled through some members actions to other members. All members in the team were allowed to take test automation tasks and learn while doing them. Team members voluntarily distributed each responsibility among themselves depending on their preference and experience level. They became more self-motivated to increase their involvement in test automation by, e.g., actively using existing test tools, fetching useful results for their needs, growing ideas that may add value for the team, and sharing the expertise with others. It was noted that test automation can be performed in a better way, when test professionals was not over-burdened with assigned tasks. 

\subsubsection{Allow time for learning curve} Allowing time for learning is a factor that determines the success of TAPI at F-Secure as well as Epics. Because of technology changes (e.g., test tools, product architecture, test infrastructure) and the discovery of new knowledge, people were expected to learn new things. Since Epics started, a learning-by-doing strategy has been applied. In the `the whole team effort model', all team members were encouraged to improve their test automation expertise and skills by performing test automation tasks. They were allowed to fail with test automation in a safe way. Experiences are discussed without scheduled meetings, sharing a team room and a discussion chat.


\subsubsection{Internal open source community mindset} Internal open source community mindset is considered critical for the success of TAPI to gain test automation maturity for continuous development. Since 2009, F-Secure has worked with internal open source community model meaning all code is visible and changeable to anyone internally. Running code served as documentation that different teams contribute on. Before 2009, projects worked on isolated branches and bringing changes together was difficult. 

In 2016, when Epics started, product code (including the test automation part) was shared in the internal open source community with other development teams, who build products from same Windows Endpoint Protection platform. Test automation assets were first created separately for Epics, and in 2018 combined with other teams on the platform into a shared repository, even if different folders. Over time, sharing assets (creating reusable methods in test automation) increased while there is still a per team separation visible in the folder structure. Shared repository with cross-team review responsibilities enabled co-creation of shared assets for test automation beyond individual team's capabilities.




\subsubsection{Test tool choice and architecture} For Epics, the use of test tools is necessary and critical for its test automation. Without the current test tools, the success of TAPI seems impossible. In 2005, the tool in use was commercial tool allowing capture and replay, used as a scripting platform. Since then, the closed languages in commercial tools were deemed limiting. 

Nowadays, the team co-owns a tailored tool set that contains more than 10 different test tools, see Fig~\ref{fig:architecture}. Each tool is used for a purpose in the present test automation system. There is many kinds of code created for test automation, for example: Nose Plugins for reusable functionalities like telemetry sending, Jenkins DSLs for job definitions in test automation, Wrappers for C++ to Python bindings, Libraries and tools for observing security incidents, and crash analyzer for post-processing analysis, scripts for specific actions and verification. Test runs and change orchestration and continuous integration related tasks are performed with Jenkins. 

Individual tools are interchangeable and got replaced when better options come along. Test tools were selected to serve current ideas, and changed as new insight emerged. Team members were allowed to identify the needs for suitable test tools. The final test tool selection decisions were made through approving a change into the system, discussing at least between two people. Each test tool was selected with experimentation mindset, to see if  that it is useful and maintainable. Benefits were discovered through experimenting with the test automation continuously running to support software production.
Integrating the variety of test tools into the same system was not a straightforward task but includes discovery. Lots of changes to the test automation system were done to combine the strengths of test tools. In the current system, test automation is based on a general purpose programming language Python that supports and integrates with the C++-based products by its design.

\subsubsection{Test lab infrastructure} ``Let us create a tool that would really enable fast provisioning of different test environments,'' said by a developer at F-Secure about ten years ago. From the insight of what could be built, with support from peers enthusiasm, they implemented a tool that framed 
the internal managed test lab infrastructure plus the images for the provisioning of various test environment. Because of the positive  results of this first attempt, since  that, this tool was put in use for a long time in product teams that are responsible for Windows endpoint security products. The provisioning system was able to start a new working test environment in 5 seconds and played a significant role in test automation progress. 

Some years after, the developer who implemented the tool left F-Secure. Future maintenance and operation was split between R\&D and IT departments - maintenance and operations with R\&D, providing infrastructure to run on from IT. With new cloud-based cost allocation, provisioning and security models, wishes to have these with the tooling emerged. In 2018, it was seen that current test lab infrastructure was under-resourced (needing more machines it got), and poorly managed (hard to find a quick fix in case of problems).  Operational problems with current system were fixed in 2019 and a future replacement is under consideration. 

\subsubsection{Product testability} 
The 2005 lessons on test automation lead to an insight on importance of product making test automation possible - testability, e.g., products requiring reboot are harder to automate, so new architecture does not require reboots. Testability was a major architectural change, not minor coordination from an outside team and test automation testers in the separate team made significant effort living with the architecture rather than changing it to testable. 

In 2009, with  a  change  in  business  focus  and  need  of solving  product  performance  issues,  the  product  architecture was completely revamped. With this change, automation testability features were designed into the products and team practices, in efforts lead by developers to ensure the feedback they need from the test automation. From intertwined functionality, the design moved towards isolated functionality in components. Information reliable automation needed was made available with C++ to Python wrappers every functionality now comes with. 

Complete redesign of product architecture enabled asking for visibility and control for test automation purposes, and getting it - or doing the necessary change yourself. Under this architecture,  services and components are  more independent. Testing a single part independently became less complex, and the effort to do it was deducted. 


\subsubsection{Telemetry}
Telemetry is process of automatically recording and transmitting the collection of data for monitoring~\cite{kechagia2015}. Product telemetry use in scale for Epics started in 2017, and expanded from product telemetry in test environments to test automation telemetry in 2019. 
Test automation telemetry solved two perceived problems: radiator snapshots were immediately outdated due to numbers of builds to test, and each failure required reading logs to know who the feedback was targeted for while being hard to collate to find trends. 

In November 2019, a Python developer in Epics implemented telemetry plugin into the test automation system for monitoring automated tests. Every automated tests automatically reported itself as it runs with telemetry. For example, it can be seen that how many automated tests are failed, passed, and skipped, how long they take. This made all of such relevant information on automated tests in scale of 200 000 tests a day visible and collated in real time. It is more straightforward than before to track and control of test automation process for continuous improvement.  

\subsubsection{Incremental approach}
\label{sec:incrementalFactor}
An emphasis was put on the incremental approach to improve test automation and its process for those 15 years in F-Secure for Windows endpoint security platform and products. From the old teams to Epics, an incremental approach was used with the point of continuous learning. Since the initial phase of TAPI, there was no documented test automation strategy specifying what should be improved for determining the mature test automation process. 
Indeed, the strategy was discovered step by step in practices. Epics continuously explored their needs and possibilities for test automation. The direction of TAPI was discussed in the groups of internal stakeholders - usually in informal settings, peer to peer. Accordingly, the actionable steps were took to make the meaningful changes. 

Rather than totally changing the whole test automation process, the changes were always added incrementally piece by piece into the existing test automation process. Some changes occurred naturally as problems arise and needed to be fixed. The change steps were carried out through experiments to allow learning, even though some succeeded and some failed. The incremental changes affected daily work of the team, both positively and negatively. The positive results became part of TAPI. 

\subsubsection{Process observation and optimization} The capability to continuously explore useful information for optimizing test automation process is critical for Epics. There was a model to specify which aspects (e.g., maintenance costs, test execution times) should be addressed for optimizing the test automation process. However, the actual optimization was done reacting to the current ideas. A principle of ``appreciate what you have and make it better" was regularly applied. Especially after `the whole team effort` model was introduced, everybody was allowed to bring or implement process optimization related ideas in the team. 



\section{Discussion}
In this section we carry out the discussion on our research question, explain the TAPI culture in F-Secure, and outline the threats to validity.

\label{sec:dicussion}
\subsection{\RQoneTitle}
In this study, we present a set of key success factors of TAPI in one DevOps team of F-Secure. We compared our results with the results of other studies in this research scope and found the similarities and differences.

Many success factors (identified in this study) have been examined in existing literature. For example, test tool choice and integration, expert team members, and self-motivated team members were mentioned by Mark Fewster and Dorothy Graham in the book `Software test automation-effective use of test execution tools', which was published in 1999\cite{fewster1999}. In other case studies (e.g.,\cite{wiklund2012, Spersson2004,grindal2006}) which were published several years ago, product testability related factors were recognized on test automation maturity studies in particular software organizations.  Our results complement the research of existing studies by showing that those factors are still valid for TAPI practices in the current industry. Nevertheless, we explore and explain new success factors, e.g., the whole team effort, incremental approach, and agile-oriented process related factors, which were rarely observed in other studies. We claim those factors should receive enough attention in future TAPI research.


On the other hand, our study results are in conflict with some observations of TAPI of the prior research: 

\begin{itemize}
    \item \textbf{Defining an explicit test automation strategy at start can guide organizations to do TAPI in general.} Test automation strategy related topics are discussed in many test maturity models like TMap~\cite{Skoomen2013tmap} and TestSPICE~3.0~\cite{TestSPICE}. Also, prior studies (e.g.,~\cite{collins2012,kasurinen2010}) surveying practitioners about the TAPI in practices confirmed that developing a test automation strategy with right concerns may contribute to the success of TAPI in Agile development environment. However, based on the observation, having a test automation strategy at the initial phase seems is not that critical for the success of TAPI in the DevOps team at F-Secure. As described in Section~\ref{sec:incrementalFactor}, there was no documented test automation strategy. With the incremental approach, there was the broad understanding about what they have now and what they want to add on the basis of existing test automation process. The goals and action plans were allowed to be discovered at any time.     
    

    \item \textbf{Selecting test tools to fit the current needs and future development.} Lots of SE researchers and practitioners have highlighted that selecting right test tools to fit the current needs and future development is critical to test automation success~\cite{raulamo2017}. However, in the observed team in this study, test tools are treated as interchangeable Lego bricks. They could be selected to serve the instant ideas, and changed as new insight emerged.
    
    \item \textbf{Measuring the quality of performance of test automation is important.} Based on the study~\cite{wang2019}, the quality of performance of test automation must be measured to reflect how the goals of TAPI are achieved. In our case, the measurements are shown, i.e., in the dashboard or telemetry. Rather than the quantitative measures,  the qualitative investigation and discussions were conducted to regular examine the changes in their TAPI. 
        
    \item \textbf{The TAPI must follow the guidelines,} as described in prior literature~\cite{furtado2014}. With the incremental approach, the observed team in this study has tailored its own test automation process depending on its requirements. 
\end{itemize}

The above differences may point out the gap between the academia and industry, though more research is needed to confirm this. 

\subsection{Test automation process improvement culture in F-Secure}

Experience report on TAPI culture with Epics reports on a company following a relaxed, verbally communicated strategy without strict rules and processes relying on developers voluntary participation. We consider this an unusual success with TAPI resulting in high maturity of test automation as well as continuously improving it. 

TAPI culture with Epics relied on the idea that a running test automation system documents itself and people working in teams and across teams in networked manner co-create continuous strategy of experimenting and improving. Appreciating what the team had, and continuously adding to it to make it better resulted in shifting the team to a place where they are happy with their automation. 

Test Automation had been team-driven to serve teams and developers as opposed to manager reporting or return on investment calculations.  Architectural layering of test automation architecture blocks, "interchangeable Lego bricks" each provide a service small enough to replace with better ideas and implementations. Customer value and TAPI value had been prioritized as equal candidates from same team effort budget. 

Finally, internal open source applied at F-Secure includes a developer-friendly sense of ownership and lack of bureaucracy to make changes towards one company's needs and goals. 

\subsection{Threats to validity}



In this section, the threats to the validity of the study in this paper and approaches taken to minimize their impacts are explored, according to a standard checklist in software engineering from Wohlin~\cite{wohlin2012}. 

\textbf{Construct validity} refers to the extent to which the study can represent the theory behind it~\cite{wohlin2012}. We reviewed prior literature about the concept of TAPI and related work before conducting the case. The study protocol was defined beforehand. We carried out several meetings to further verify the content of experience reports and complement the details. At the end, our observations and study results were reviewed and verified with the representatives (who involved in TAPI in Epics) to avoid false interpretations and ensure the reliability.

\textbf{External validity} is concerned with how the study results can be generalized~\cite{wohlin2012}. The study setting of this paper may threaten the external validity. Our findings are strongly bounded by the context of this team at F-Secure, and they may not be representative for TAPI of all software organizations. To address this threat, we attempted to describe the company and the team in as much detail as possible, but since the time span of their TAPI is large it is hard to acquire detailed information about the context of F-Secure conducting TAPI in very early years. This makes it more challenge to relate the case in this paper to other similar TAPI cases in the industry. Individual differences are suggested to be considered when generalizing the study results.

\textbf{Conclusion validity} refers to whether the correct conclusions are made through observations of the study~\cite{wohlin2012}. In our study, the conclusions were made according to the thematic analysis on raw data. We performed the data analysis  with NVivo in where all qualitative codes were stored. The conclusions were verified among co-authors.  
 
\textbf{Internal validity} focus on how the study causes the outcomes~\cite{wohlin2012}.  In our study, threats to internal validity may lie in the data collection. Our data were mainly collected from a collection of experience notes, team reflection reports, and telemetry result reports. Materials were provided by one person (the second author of this paper) in Epics at F-Secure. They were filtered through models available in literature practitioners suggest may provide an outdated perspective to maturity. Viewpoints on the material as well as selecting aspects to highlight from the material might have  some subjective viewpoints. We tried to overcome this type of threat by acquiring more quantitative data to explain the results. However, because of the personnel and technology changes in passed years, collecting and comparing quantitative data beyond what was accessible at first was outside scope of this work. 


\section{Conclusion}
\label{sec:conclusion}

This paper presents an experience report on TAPI in one DevOps team at F-Secure. For the study purpose, we reviewed a collection of experience notes, team reflection reports, and telemetry result reports. Several meetings were held to discuss the details. As the study results, first, we reported that, the team defined its test automation maturity for continuous development by a set of indicators, see Section~\ref{sec:casedescription}. Second, it is noted that, to successfully conduct TAPI, the team has performed main practices around a set of factors mapped into different dimensions, see Table~\ref{tab:factors}. Third, under the further investigation, we found that the team has the tailored test automation process for continuous development, which may have the similarities or differences with the ones defined in prior literature.   


This study has three main contributions. First, from the industry perspective, it introduces the industrial case of successfully carrying out TAPI in a DevOps team. Second, from the academia perspective, this study connects to the prior studies and makes novel contribution. For example, the success factors of TAPI frequently mentioned in prior studies are explained with empirical evidence. Also, we identified new factors such as the whole team effort, incremental approach, and telemetry. Third, as an empirical research, this study narrows the gap between academia and industry. 

In the future, we plan to assess the level of test automation maturity in the same DevOps team at F-Secure. Based on the assessment results, we could investigate the specific impact of critical success factors on the maturity level and the short-term and long-term benefits and effects of test automation. Additionally, the set of key success factors presented in this paper may be only a part of possible solutions to some software organizations. We also could widen the research by carrying out a case study surveying more software organizations.

\section*{Acknowledgment}
The first and third author of this study are supported by TESTOMAT Project (ITEA3 ID number 16032), funded by Business Finland under Grant Decision ID 3192/31/2017.

\bibliographystyle{IEEEtran}
\bibliography{ownbib}

\end{document}